\documentstyle[12pt]{article}

\begin{document}

\title{Does the Wave Function Provide a Complete Description of Physical
Reality?}

\bigskip
\author{Sheldon Goldstein\ and Joel L. Lebowitz\\
Departments of Mathematics and Physics, Rutgers University\\ New Brunswick,
NJ 08903}

\maketitle

Quantum mechanics is undoubtedly the most successful theory yet devised by
the human mind.  Not one of the multitude of its calculated predictions has
ever been found wanting, even in the last measured decimal place---nor is
there any reason to believe that this will change in the foreseeable future.
All the same, it is a bizarre theory.  Let us quote  Feynman
\cite{Fey}, one
of the deepest scientist-thinkers of our century and one not known for his
intellectual (or any other) modesty, on the subject: ``There was a time
when the newspapers said that only twelve men understood the theory of
relativity.  I do not believe there ever was such a time.  There might have
been a time when only one man did, because he was the only guy who caught
on, before he wrote his paper.  But after people read the paper a lot of
people understood the theory of relativity in some way or other, certainly
more than twelve.  On the other hand, I think I can safely say that nobody
understands quantum mechanics. ... I am going to tell you what nature
behaves like.  If you will simply admit that maybe she does behave like
this, you will find her a delightful, entrancing thing.  Do not keep saying
to yourself, if you can possibly avoid it, `but how can it be like that?'
because you will get `down the drain,' into a blind alley from which nobody
has yet escaped.  Nobody knows how it can be like that.''

Feynman's point of view, expressed as usual with great vigor and clarity,
characterizes the attitude of most physicists towards the foundations of
quantum mechanics, a subject concerned with the meaning and interpretation
of quantum theory---at least it did so before the work of  Bell
\cite{Bellepr} and the experiments of Aspect et al. [\cite{Aspect}:b]
(which came after the cited Feynman lecture).  Even today, the subject is
treated like a poor stepchild of the physics family.  It is pretty much
ignored in most standard graduate texts, and what is conveyed there is
often so mired in misconception and confusion that it usually does more
harm than good.  To many physicists it appears that not only does
foundational research not lead to genuine scientific progress, but that it
is in fact dangerous, with the potential for getting people ``down the
drain.'' Even to the more tolerant ones it often seems that what is
achieved merely supports what every good physicist should have known
already.

While these perceptions are partly true, they are also in part
misapprehensions arising from the, to our taste, much too practical approach
taken by many physicists.  Basic questions concerning the physical meaning
of  quantities, such as the wave function, which we manipulate in our
computations are too important to be left to philosophers. One such
question, whether the description of a physical system provided by
its wave function is complete, is central to the articles reprinted in this
section. [We refer to papers in this book by [ :b] and to papers on the
CD-ROM by [ :c].]

Einstein, Podolsky, and Rosen (EPR) [\cite{EPR}:b] argue that quantum
mechanics provides at best an incomplete description of physical reality.
Indeed, they claim that there are situations in which the very predictions
of quantum theory demand that there be elements of physical reality---i.e.,
predetermined, preexisting values for physical quantities, which are {\it
revealed\/} rather than {\it created\/} if and when we measure those
quantities---that are not incorporated within the orthodox quantum
framework.  In the original version of the argument, these elements of
reality are the (simultaneous) values of the position and momentum of a
particle belonging to an EPR pair---a pair of particles whose quantum
state, given by the EPR wave function, involves such strong quantum pair
correlations that the position or momentum of one of the particles can be
inferred from the measurement of that of the other. By the uncertainty
principle, however, the position and momentum of one particle cannot
simultaneously be part of the quantum description. In the later version of
the EPR analysis due to Bohm \cite{Bohmqt}, which provides the
framework for most of the experimental tests of quantum theory that were
stimulated by the celebrated Bell's inequality paper \cite{Bellepr}, these
elements of reality are the values of the (simultaneous) components, in all
possible directions, of the spins of the particles belonging to a Bohm-EPR
pair---a pair of spin 1/2 particles prepared in the singlet $S=0$
state---or, in another version, the simultaneous components of photon
polarization in a suitable photon pair.  We shall call these the Bohm-EPR
elements of reality. (They again cannot simultaneously be part of the
quantum description because spin components in different directions do not
commute.)

The EPR analysis begins with a criterion of reality: {\it ``If, without in
any way disturbing a system, we can predict with certainty ... the value of a
physical quantity, then there exists an element of physical reality
corresponding to this physical quantity.''\/} EPR continue, ``It seems to
us that this criterion, while far from exhausting all possible ways of
recognizing a physical reality, at least provides us with one such
way .... Regarded not as a necessary, but merely as a sufficient, condition of
reality, this criterion is in agreement with classical as well as
quantum-mechanical ideas of reality.''  They then deduce the existence of
the relevant elements of reality for an EPR pair from the predictions of
quantum theory for the pair. In so doing, however, they crucially require a
locality assumption, that ``the process of measurement carried out on the
first system ... does not disturb the second system in any way.''  EPR
conclude as follows: ``While we have thus shown that the wave function does
not provide a complete description of the physical reality, we left open
the question of whether or not such a description exists. We believe,
however, that such a theory is possible.''

We wish to emphasize that in arguing here for the incompleteness of the
quantum description, EPR were not questioning the validity of the
experimental predictions of quantum theory. On the contrary, they were
claiming that these predictions were not only compatible with a more
complete description---in particular, one involving their elements of
reality---but also demanded one. Elsewhere, Einstein \cite{Schilpp}
asserts that in ``a complete physical description, the statistical quantum
theory would ... take an approximately analogous position to the statistical
mechanics within the framework of classical mechanics.''

Niels Bohr [\cite{Bohrb}:b], in what is perhaps the definitive statement of his
principle of complementarity, disagreed with the EPR conclusion, though he
did not take the EPR analysis lightly. The central objection in Bohr's
reply is that the EPR reality criterion ``contains an ambiguity as regards
the meaning of the expression `without in any way disturbing a system.' Of
course, there is ... no question of a mechanical disturbance .... But ...
there is essentially the question of an influence on the very conditions
which ... constitute an inherent element of the description of any
phenomenon to which the term `physical reality' can be properly attached
....'' While, with Bell \cite{Bellbook}, we ``have very little idea what
this means,'' it does perhaps suggest ``the feature of wholeness typical of
proper quantum phenomena'' elsewhere stressed by Bohr \cite{Bohr}.

Bohm [\cite{Bohmb}:b], [\cite{Bohmc}:c], on the other hand, not only
agreed with EPR that the quantum description is incomplete, but showed
explicitly how to extend the incomplete quantum description---by the
introduction of ``hidden variables''---into a complete one, in such a way
that the indeterminism of quantum theory is completely eliminated. We shall
call Bohm's deterministic completion of nonrelativistic quantum theory
Bohmian mechanics. In Bohmian mechanics the ``hidden variables'' are simply
the positions of the particles, which move, under an evolution governed by
the wave function, in what is in effect the simplest possible manner
\cite{DGZ}. We should emphasize that Bohmian mechanics is indeed an
extension of quantum theory, in the sense that in this theory, as in
quantum theory, the wave function evolves autonomously according to
Schr\"odinger's equation. Moreover, it can be shown \cite{DGZ} that the
statistical description in quantum theory, given by $\rho=|\psi|^2$, indeed
takes, as Einstein wanted, ``an approximately analogous position to the
statistical mechanics within the framework of classical mechanics.''

Bohmian mechanics was ignored by most physicists, but it was taken very
seriously by Bell, who declared \cite{Bellpw} that ``in 1952 I saw the
impossible done.''  Bell quite naturally asked how Bohm had managed to do
what von Neumann \cite{vN} had proclaimed to be---and almost all
authorities agreed was---impossible. (It is perhaps worth noting that
despite the almost universal acceptance among physicists of the soundness
of von Neumann's proof of the impossibility of hidden variables,
undoubtedly based in part on von Neumann's well-deserved reputation as one
of the greatest mathematicians of the twentieth century, Bell \cite{M}
felt that the assumptions made by von Neumann about the requirements for a
hidden-variable theory are so unreasonable that ``the proof of von Neumann
is not merely false but {\it foolish!''\/} See also Ref. 16.) His
ensuing hidden-variables analysis led to  Bell's inequality, which
must be satisfied by certain correlations between Bohm-EPR elements of
reality---and, of course, by correlations between their measured values.
He observed also that quantum theory predicts a sharp violation of the
inequality when the quantities in question are measured.

Thus the specific elements of reality to which the EPR analysis would lead
(if applied to the Bohm-EPR version) must satisfy correlations that are
incompatible with those given by quantum theory. That is, these elements of
reality, whatever else they may be, are demonstrably incompatible with the
predictions of quantum theory and hence are certainly not part of any
completion of it. It follows that there is definitely something wrong with
the EPR analysis, since quantum mechanics cannot be (even partially)
completed in the manner demanded by this analysis. In other words, had EPR
been aware of the work of Bell, they might well have predicted that quantum
theory is wrong and proposed an experimental test of Bell's inequality to
settle the issue once and for all.

Of course, EPR were not aware of Bell's analysis, but Clauser, Horne,
Shimony, and Holt were [\cite{CHSH}:c]. Their proposal for an experimental
test has led to an enormous proliferation of experiments, the most
conclusive of which was perhaps that of Aspect et al.  [\cite{Aspect}:b]
included here.  The result: Quantum mechanics is right.

We note, however, that the predictions of (nonrelativistic) quantum
mecha\-nics---in particular, those for the experimental tests of Bell's
inequality---are in complete agreement with the predictions of Bohmian
mechanics. Thus the Bohm-EPR elements of reality are not part of Bohmian
mechanics! This is because in Bohmian mechanics the result of what we speak
of as measuring a spin component depends as much upon the detailed
experimental arrangement for performing the measurement as it does upon
anything existing prior to and independent of the measurement. This
dependence is an example of what the experts in the hidden-variables field
call contextuality \cite{Bellrmp}, (see also Ref. 15) i.e.,
of the critical importance of not overlooking ``the interaction with the
measuring instruments which serve to define the conditions under which the
phenomena appear''\cite{BE}.

In fact, just before he arrived at his inequality, Bell noticed that ``in
this [Bohm's] theory an explicit causal mechanism exists whereby the
disposition of one piece of apparatus affects the results obtained with a
distant piece. ...  Bohm of course was well aware of these features of his
scheme, and has given them much attention.  However, it must be stressed
that, to the present writer's knowledge, there is no {\it proof\/} that
{\it any\/} hidden variable account of quantum mechanics {\it must\/} have
this extraordinary character. It would therefore be interesting, perhaps,
to pursue some further `impossibility proofs,' replacing the arbitrary
axioms objected to above by some condition of locality, or of separability
of distant systems''\cite{Bellrmp}. Almost immediately, Bell found his
inequality.  Thus did Bohmian mechanics lead to Bell's refutation of the
EPR claim to have ``shown that the wave function does not provide a
complete description.''  At the same time it showed, by explicit example,
the correctness of the EPR belief ``that such a theory is possible''!

While Bell's analysis, together with the results of experiments such as
Aspect's, implies that the EPR analysis was faulty, where in fact did EPR
go wrong? Since their only genuine assumption was that of locality quoted
above, and since their subsequent reasoning is valid, it is this assumption
that must fail, both for quantum theory and for nature herself.  Aspect's
experiment thus establishes perhaps the most striking implication of
quantum theory: Nature is nonlocal!  This conclusion is of course implicit
in the very structure of quantum theory itself, based as it is on a
field---the wave function---which for a many-body system lives not on
physical space but on a 3$n$-dimensional configuration space, a structure
that allows for the entanglement of states of distant systems---as most
dramatically realized in the EPR state itself. But while quantum mechanics
may someday be replaced by a theory of an entirely different character, we
may nonetheless conclude---though there are some who disagree
\cite{M}---from Bell and Aspect that the nonlocality it implies is here
to stay.

One of the great foundational mysteries that remains very much unsolved is
how nonlocality can be rendered compatible with special relativity, i.e.,
with Lorentz invariance. Here Bohmian mechanics is of no direct help, since
it manifestly and fundamentally is not Lorentz invariant. But there is no
reason to believe that a more appropriate completion of quantum theory, one
that is Lorentz invariant and perhaps even generally covariant, cannot be
found. However, one should not expect finding it to be easy.

One lesson of this story is perhaps that we would be wise to place greater
trust in the mathematical structure of quantum theory, and less in the
philosophy with which quantum theory is so often encumbered. For the EPR
problem, the mathematical structure correctly suggests nonlocality, while
the philosophy makes the questionable demand that the wave function provide
a complete description, at least on the microscopic level.  The paper
[\cite{AB}:b] by Aharonov and Bohm included here supports this
lesson. Aharonov and Bohm dramatically demonstrate that the electromagnetic
vector potential has a reality in quantum theory far beyond what it has
classically: a nonvanishing vector potential may generate a shift in an
interference pattern for an electron confined to a region in which the
magnetic field itself vanishes.  The Aharonov-Bohm effect, while rather
clear from the role played by the vector potential in Schr\"odinger's
equation, is rather surprising from the perspective of the usual quantum
philosophy, which, in attempting to explain quantum deviations from
classical behavior, appeals to limitations on what can be measured or known
arising from disturbances occurring during the act of measurement that are
due to the finiteness of the quantum of action.

It is appropriate to mention at this time---even though it is not the focus
of any of the five papers included in this chapter---one of the strongest
arguments for the conclusion that the quantum mechanical description is
incomplete: the notorious measurement problem---or, what amounts to the
same thing, the paradox of Schr\"odinger's cat. The problem is that the
after-measurement wave function for system and apparatus arising from
Schr\"odinger's equation for the composite system typically involves a
superposition over terms corresponding to what we would like to regard as
the various possible results of the measurement---e.g., different pointer
orientations. Since it seems rather important that the actual result of the
measurement be a part of the description of the after-measurement
situation, it is difficult to see how this wave function could be the
complete description of this situation.  By contrast, with a theory or
interpretation in which the description of the after-measurement situation
includes, in addition to the wave function, at least the values of the
variables that register the result, the measurement problem vanishes. (The
remaining problem of then justifying the use of the ``collapsed'' wave
function---corresponding to the actual result---in place of the original
one is often confused with the measurement problem. The justification for
this replacement is nowadays frequently expressed in terms of decoherence.
One of the best descriptions of the mechanisms of decoherence, though not
the word itself, can be found in the Bohm article reprinted here; see also
Ref. 5.  We wish to emphasize, however, as did Bell in his article
``Against Measurement'' \cite{AM}, that decoherence per se in no way comes
to grips with the measurement problem itself.)

The orthodox response to the measurement problem is that we must
distinguish between closed systems and open systems---those upon which an
external ``observer'' intervenes. While we do not want to delve into the
merits of this response here---nor is this the place to discuss the sundry
proposals for alternate interpretations of quantum theory, such as those of
Schulman \cite{Schulman} and of Pearle [\cite{Pearlec}:c], \cite{P} and
Ghirardi, Rimini, and Weber \cite{GRW}, \cite{Bellj}---we do wish to note
one particular difficulty, much emphasized of late.  This concerns the
now-popular subject of quantum cosmology, concerned with the physics of the
universe as a whole, certainly a closed system! A formulation of quantum
mechanics that makes sense for closed systems seems to be demanded.
Bohmian mechanics is one such formulation.  Others also now generating a
good deal of excitement are due to Griffiths \cite{G},
Omn\`es \cite{O}, and  Gell-Mann and  Hartle \cite{GMH}.  All
of these exemplify the EPR conclusion ``that the wave function does not
provide a complete description of the physical reality.''

\end{document}